\documentclass[usenatbib]{mn2e}
\usepackage{amsmath}
\usepackage{graphicx}
\usepackage{natbib}
\def\apj{{ApJ}}                 
\def\apjl{{ApJ}}                
\def\apjs{{ApJS}}               
\def\mnras{{MNRAS}}             
\def\prd{{Phys.~Rev.~D}}        
\def\pasj{{PASJ}}               

\def\physrep{{Phys.~Rep.}}   

\newcommand{\mpc}{\rm {h^{-1}Mpc }}

\newcommand{\avg}[1]{\langle{#1}\rangle}

\newcommand{\ltsima}{$\; \buildrel < \over \sim \;$}
\newcommand{\lsim}{\lower.5ex\hbox{\ltsima}}
\newcommand{\gtsima}{$\; \buildrel > \over \sim \;$}
\newcommand{\gsim}{\lower.5ex\hbox{\gtsima}}

\def\gtrsim{\mathrel{\hbox{\rlap{\hbox{\lower4pt\hbox{$\sim$}}}\hbox{$>$}}}}

\def\lesssim{\mathrel{\hbox{\rlap{\hbox{\lower4pt\hbox{$\sim$}}}\hbox{$<$}}}}

\title[Monopole of the three point correlation function of 2dF]{
The Monopole Moment of the Three-Point Correlation 
Function of the 2-degree Field Galaxy Redshift Survey}

\begin{document}
 
\author[Pan \& Szapudi]{Jun Pan$^{1,2}$, Istv\'an Szapudi$^{1}$\\
$^{1}$Institute for Astronomy, University of Hawaii,
2680 Woodlawn Dr, Honolulu, HI 96822, USA \\
$^{2}$School of Physics and Astronomy, University of Nottingham, 
Nottingham NG7 2RD, UK
}
\maketitle

\begin{abstract}

We measure the monopole moment of the three-point correlation
function on scales $1\mpc-70\mpc$ in the 
Two degree Field Galaxy Redshift Survey (2dFGRS).
Volume limited samples are constructed using a series of
integral magnitudes bins between $M = -18 \ldots -22$. Our measurements
with a novel edge-corrected estimator represent
most, if not all, three-point level monopole or angular averaged
information in the catalogue. We fit a perturbative
non-linear bias model to a joint data vector formed from
the estimated two- and three-point correlation functions.
Two different models are used:
an analytic model based on Eulerian perturbation 
theory including bias and redshift distortions, and
a phenomenological bias model based on the direct redshift space measurements
in the large Virgo simulations. To interpret the clustering results,
we perform a three parameter Gaussian
maximum likelihood analysis. In the canonical $-21 \sim -20$ volume
limited sample we find $\sigma_8 =  0.93^{+0.06}_{-0.2}$,
$b = 1.04^{+0.23}_{-0.09}$, and $b_2 = -0.06^{+0.003}_{-0.001}$.
Our estimate of $\sigma_8$, is robust across the different
volume limited samples constructed. These results, based
solely on the large scale clustering of galaxies,  are 
in excellent agreement with
previous analyses using the Wilkinson Anisotropy Probe:
this is a spectacular success of the concordance model. 
We also present two-parameter fits for
the bias parameters, which are in excellent agreement
with previous findings of the bias evolution in the 2dFGRS.

\end{abstract}

\begin{keywords}
large scale structure --- cosmology: theory --- methods:
statistical
\end{keywords}

\section{Introduction}

Statistical analyses of the Two Degree Field Galaxy
Redshift Survey (2dFGRS) \citep{CollessEtal2001}
have propelled significant progress in high precision cosmology. 
For instance, 
measurements of the two-point correlation 
function and the power spectrum have provided tight constraints
on the theories of structure formation 
\citep[e.g.][]{PercivalEtal2001, NorbergEtal2001, 
HawkinsEtal2003, ColeEtal2005}. The large volume and
high quality of the 2dFGRS encourage further studies
of higher order statistics. Such investigations provide
information on the Gaussianity of the small initial dark 
matter fluctuations,
the emergence of non-Gaussianity through non-linear gravitational
effects, and even the murky physical processes of galaxy formation.
The latter might manifest itself as ``bias'' \citep{Kaiser1984},
where the clustering of galaxies might be statistically different
from that of the dark matter. Higher than second order statistics
provide the only tool with which to separate these effects from
the gravitational amplification and initial conditions. 
The resulting constraints on the bias are
interesting in their own right,  and they provide the new avenues to
ultimately constrain cosmological parameters.

Third order statistics represent the first non-trivial step in
the perturbative understanding of non-Gaussianity.
Indeed, to date numerous works have been devoted mainly 
to the measurement and understanding of the third order 
statistics of 2dFGRS. \citet{VerdeEtal2002}
estimated bias parameters from their bispectrum measurement
at wave lengths $0.1 < k < 0.5{\rm h/Mpc}$. Three point correlation
function $\zeta$ of an early released 2dF sample (2dF100k) is measured 
by \citet{JingBorner2004} focusing on empirical
formula fit for $\zeta$. \citet{WangEtal2004} measured
three point correlation function for the 2dFGRS on small scales
to test their conditional luminosity function model
jointly with the halo model. As an alternative to the three-point
correlation function or bispectrum,
\citet{CrotonEtal2004} calculated moments of counts
in cells, or averaged $N$-point correlation 
functions, on $1-9\mpc$ scales, and estimated relative
bias parameters.

A common thread in previous measurements was that they focused
on either relatively small scales and/or a set of hand picked subset of 
triangular configurations, which characterize three-point statistics.  
With the most natural parametrizations predominantly used in the past, such as
the three sides of a triangle $\zeta(r_1, r_2, r_3)$, or
$\zeta(r_1, r_2, \theta)$ with $\theta$ the angle between $r_1$ and $r_2$,
it is both computationally burdensome, and conceptually
difficult to measure and interpret three-point statistics
in a large dynamic range. Recently, \citet{Szapudi2004a} has shown
that a multipole expansion motivated by rotational invariance
helps substantially with this ``combinatorial expansion'' of parameters. 
It was demonstrated that it is most efficient to 
expand spherically symmetric functions of two unit vectors into
$L=0$ bipolar spherical harmonics. 
In turn, the recipe boils down to expand $\zeta$
into Legendre polynomials $P_\ell(\cos\theta)$ 
\begin{equation}
\zeta(r_1, r_2, \theta) 
= \sum_{\ell=0}^\infty \frac{2\ell+1}{4 \pi} 
\zeta_\ell(r_1, r_2) P_\ell(\cos \theta) 
\end{equation}
\citet{Szapudi2004a} has demonstrated that the first few multipoles,
often up to $\ell=2$, concentrate most of the useful three-point level 
information. While one
still needs to consider the scale dependence over $r_1$ and $r_2$,
the configuration space effectively becomes two-dimensional.

In addition to the conceptual simplification, the above parametrization
suggests new algorithms to calculate three-point functions 
\citep{Szapudi2005b}. 
In particular, \citet{PanSzapudi2005} adapted the
fully edge corrected estimator of
\citet{SzapudiSzalay1998} for the $\ell=0$ monopole moment,
and have demonstrated a simple and fast $N^2$ algorithm for its
calculation.  Edge correction is a major advantage over other
three-point statistics which can be calculated reasonably fast, 
most notably the bispectrum and moments of counts in cells. 
This means that the resulting estimates of the monopole of the
three-point function are expected to be
more robust against complicated geometry of the window, 
cut out holes etc, than other previously used measures. 
Since real surveys, such as the 2dFGRS have complicated
spatial structure, edge effect correction is a must when approaching
large scales.

The monopole moment captures all information about the amplitude
of the three-point correlation function; all other multipoles provide
information on the shape.
Besides the fact  that the monopole moment is the lowest order,
thus the simplest to measure and interpret in the multipole series, it 
also has a simple transformation under bias,
and we have a relatively
accurate understanding of its redshift distortions 
\citep{PanSzapudi2005}. These properties single out the monopole
moment as a principal candidate among the three-point statistics
for practical applications.

In this paper we set out to harvest the fruits of recent
theoretical developments, and measure and interpret 
the monopole moment of the three-point function in the 2dFGRS. 
The interpretation
of three-point statistics in terms of bias parameters was
put forward by \citet{Fry1994}. His method was later
perfected to include maximum likelihood fits and more sophisticated
theoretical-numerical modeling of ratio statistics
\citep{MatarreseEtal1997,VerdeEtal1998,Scoccimarro2000,FeldmanEtal2001,VerdeEtal2002,GaztanagaScoccimarro2005}.
We develop a novel joint maximum likelihood technique using both
two- and three-point statistics (as opposed to ratio statistics) 
for simultaneous estimation of
bias coefficients and cosmological parameters, such as $\sigma_8$.
We estimate covariances in the data using mock surveys,
and constrain the parameters of our theory in 
a Gaussian maximum likelihood context with
scales up to $140\mpc$ entering into the analysis.
Even though  $\zeta_0$ is only the first in the series
of multipoles, we will see that it contains invaluable,
hitherto untouched information on cosmology and bias.

The next section outlines our method of estimating $\zeta_0$
in volume limited subsamples of 2dFGRS;
section 3 details the theoretical framework for the 
interpretation of the data in terms of bias and cosmological
parameters; the resulting constraints are presented
in section 4; discussion and summary follows in section 5.

\section{Measurement of $\zeta_0$}

\subsection{The data set}

In order to estimate three-point correlation functions, we constructed
volume limited samples from  the 2dFGRS final 
data release spectroscopic catalogue \citep[the 2dF230k,][]{CollessEtal2003}
with 221414 galaxies of good redshift quality $Q\ge3$ 
\citep{CollessEtal2001}.
Excluding the ancillary random fields leaves us two large contiguous
volumes: one near the South Galactic Pole (SGP) covering approximately
$-37^\circ\negthinspace.5<\delta<-22^\circ\negthinspace.5$, $21^{\rm
h}40^{\rm m}<\alpha<3^{\rm h}40^{\rm m}$ and the other one around
the North Galactic Pole (NGP) defined roughly by
$-7^\circ\negthinspace.5<\delta<2^\circ\negthinspace.5$, $9^{\rm
h}50^{\rm m}<\alpha<14^{\rm h}50^{\rm m}$. 
The parent sample was further restricted
by completeness $f>0.7$, and apparent magnitudes limits in photometric 
$b_J$ band with bright cut of $m_{b_J}=15$ and faint cut of
median value of $\sim19.3$ with certain small variation as specified by
masks \citep{CollessEtal2001}. 
 
Volume limited subsamples are built from the parent sample by selecting 
galaxies in specified absolute magnitude ranges. 
These were calculated with
$k+e$ correction as in \citet{NorbergEtal2002}. The most important
properties of the result SGP and NGP
samples are summarized in Table~\ref{tb:2dfdata}.
For our measurements, the NGP and SGP were combined together to
achieve the highest possible volume. 

\begin{table*}
\centering
\caption{Volume limited subsamples of 2dFGRS.
Comoving distance $d$ is calculated from
redshifts $z$ assuming $\Lambda$CDM universe 
with $\Omega_\Lambda=0.7$, $\Omega_m=0.3$.}
 
\begin{tabular}{lccccccc}
\hline
$M_{b_J}-5\log_{10} h$ & $z_{min}$ & $z_{max}$ & $d_{min}$ & $d_{max}$ & $N_g^{SGP}/N_g^{NGP}$ &
 $\bar n(10^{-3}h^3{\rm Mpc}^{-3})$ \\
\hline
-18 --- -17 & 0.0131 & 0.0575 & 39.0 & 170.0 & 4046/3192    & 12.97\\
-19 --- -18 & 0.0205 & 0.087  & 61.2 & 255.7 & 11935/9625   & 11.35\\
-20 --- -19 & 0.0320 & 0.129  & 95.2 & 374.9 & 23595/17729  & 6.922\\
-21 --- -20 & 0.0495 & 0.186  & 146.6 & 532.9 & 18081/12499 & 1.798\\
-22 --- -21 & 0.0754 & 0.261  & 222.2 & 735.7 & 4095/2113   & 0.140\\
\hline
\end{tabular}
\label{tb:2dfdata}

\end{table*}

\subsection{Estimation of the correlation function}

Two point correlation functions are
measured with the \citet{LandySzalay1993} estimator
\begin{equation}
\hat{\xi}=\frac{DD-2DR+RR}{RR}\ .
\end{equation}
Here $D$ stands for data and $R$ for points selected from
random catalogues. These were created according to the exact
geometry and completeness masks of the subsamples,
with 20 times as many random points as the number of galaxies $N_g$.

The three point correlation function also has similar minimum
variance estimator
\citep{SzapudiSzalay1998}
\begin{equation}
\hat{\zeta}=\frac{DDD-3DDR+3DRR-RRR}{RRR}\ .
\label{eq:ss}
\end{equation}
This estimator has been shown by \citet{KayoEtal2004} to be
more accurate than other estimators currently in use.
Because of the volume limited samples, no additional weighting
was necessary. 

For the present analyses, we measured the monopole moment
or angular average of the three-point
correlation function \citep{Szapudi2004a}
\begin{equation}
\zeta_0(r_1,r_2) = 2\pi\int d\cos\theta\zeta(r_1,r_2,\theta)\ .
\end{equation}
Since this is an angular averaged quantity, the estimator of Eq.~\ref{eq:ss} 
can be realized with setting up bins of shells around center points.
These are essentially {\em neighbour counts in shells}, and
can be realized with a simple $N^2$ estimator put forward in
\citet{PanSzapudi2005}.
Explicitly, if the number of galaxies around galaxy $i$ 
in bin $(r_1^{lo}, r_1^{hi})$ is $n_i(r_1)$, 
in bin $(r_2^{lo}, r_2^{hi})$ is $n_i(r_2)$,
the $DDD$ in Eq.~\ref{eq:ss} reads
\begin{equation}
DDD=\left\{ \begin{array}{cc} \frac{\sum_{i=1}^{N_g} n_i(r_1) n_i(r_2)}{N_g (N_g
-1) (N_g-2)}, & if \, r_1 \ne r_2 \\
\frac{\sum_{i=1}^{N_g} n_i(r_1) \left(n_i(r_2)-1\right)}{N_g (N_g-1) (N_g-2)}\ ,
 & if \, r_1 = r_2 \end{array} \right.\ .
\end{equation}
This estimator has no Poisson noise bias
as there is no overlap between the configurations.
The lack of shot noise bias, and the precise edge correction are major
technical advantages over measuring the bispectrum 
via direct Fourier transform \citep[e.g.][]{Scoccimarro2000,VerdeEtal2002},
or moments of counts in cells
\citep[e.g.,][and references therein]{SzapudiColombi1996, Szapudi1998a, BernardeauEtal2002}.

Redshift space measurements were performed in $29$ logarithmic bins between
$0.7-70\mpc$ for the two-point correlation function, and all corresponding
pairs for $\zeta_0(r_1,r_2)$, altogether $29\times 30/2=435$ bins.
All measurements were repeated in the available $22$ mock 2dF surveys,
as well as eight equal sub cubes selected from the
Virgo VLS $\Lambda$CDM simulations \citep{MacfarlandEtal1998}.
 
\begin{figure}
\resizebox{\hsize}{!}{
\includegraphics{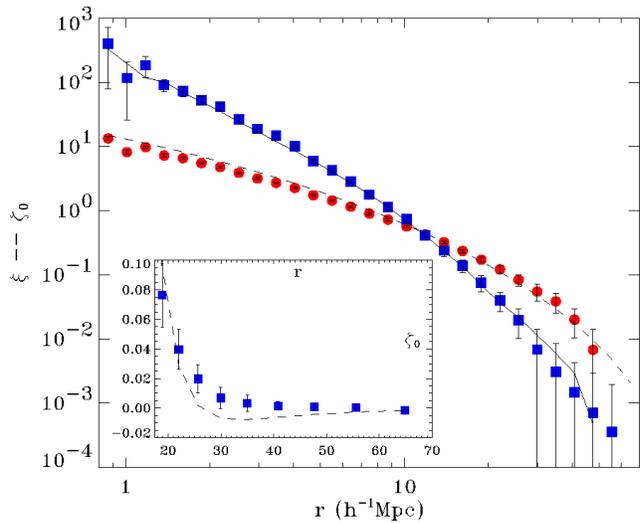}
}
\caption{ We display our measurements of the two- and three-point
correlation functions $\xi$ (red solid circles) and $\zeta_0$ 
(blue solid squares) in the canonical M=$-21 \sim -20$ volume limited sample.
Error bars have been calculated from 22 2dF mocks. Solid line is our
phenomenological model with parameters in Table~\ref{table:ml3}, dashed lines
are theoretical model with parameters in Table~\ref{table:ml2t}. 
Inset plotted on
a linear scale for $\zeta_0$ on large scales.
}
\label{fig:xi-zeta}
\end{figure}

\section{Interpretation}

\subsection{Theoretical Framework}

To interpret the clustering present in the 2dF, we use two models:
theoretical model for large scales, and
a phenomenological model for intermediate scales. 

The theoretical model uses Eulerian perturbation theory 
\citep[e.g.,][and references therein]{BernardeauEtal2002}
to calculate the real space dark matter three-point correlation
function; specifically, we use the formulae in \citet{Szapudi2005a}.

Our bias model is motivated by the usual perturbative expansion
$\delta_g = b\delta+b_2\delta/2+\ldots$ \citep{FryGaztanaga1993}
\begin{equation}
\begin{aligned}
\xi_g &= f_2 b^2\left(\frac{\sigma_8}{0.9}\right)^2\xi\\
 \zeta_{g,0} &= \left(\frac{\sigma_8}{0.9}\right)^4
 \left[b^3f_3\zeta_0+b_2 b^2f_2^2(\xi_1\xi_2+\xi_2\xi_3+\xi_3\xi_1)_0\right],
\end{aligned}
\label{eq:model}
\end{equation}
where $b$ and $b_2$ are the linear and non-linear bias factors,
$(...)_0$ denotes angular averaging,
and $f_2,\, f_3$ are redshift distortion enhancement factors
\citep{Kaiser1987, Hamilton1998}
\begin{equation}
\begin{aligned}
    f_2  &=  \left( 1+\frac{2}{3}f+\frac{1}{5}f^2 \right)\\
    f_3 &= \frac{ 5(2520 + 3360 f+1260 f^2+9 f^3-14 f^4)}{98(15+10f+3f^2)^2} 
    \cdot\frac{7}{4}\ f_2^2,
\end{aligned}
\label{eq:zdist}
\end{equation}
where $f = \Omega^0.6/b$.
The third order $f_3$ is obtained by \citet{PanSzapudi2005}
through angle averaging perturbation theory results of
\citet{ScoccimarroEtal1999}.
\citet{PanSzapudi2005} have shown this this is a good approximation
to angle averaged three-point quantities with deviation at $5\%$ level.

Our phenomenological model is obtained by using directly
our measured Virgo VLS redshift space two and three-point
correlation functions in Equation~\ref{eq:model} (with
$f_2=f_3=1$ formally), with the exception that we still
use the theory for $f_2^2(\xi_1\xi_2+\ldots)_0$ term.

\subsection{Data Vectors}

We intend to analyse the data in the above theoretical framework
using maximum likelihood methods. When constructing a data
vector from the measurements, we opt for not using
ratio statistics of the sort $Q_3\simeq \zeta/\xi^2$
as it has been done in all previous studies, 
since our measurements include such large scales 
where $\xi$ becomes an extremely
small number, possibly crosses zero. 
Our choice has the additional advantage that no ratio bias
\citep[e.g.,][]{SzapudiEtal1999} appears, and our data
have direct dependence on $\sigma_8$ which we will exploit.

We construct the data vector $(\xi,\zeta_0)$
in an appropriate scale range, where the theoretical model
is expected to be a good approximation (c.f. Table~\ref{table:scales}).
Note that we excise redundant
entries from the $\zeta_0(r_1,r_2)$ matrix, i.e. each pair
of scales appears only once.

\subsection{Covariance Matrix}

\begin{figure}
\resizebox{\hsize}{!}{
\includegraphics{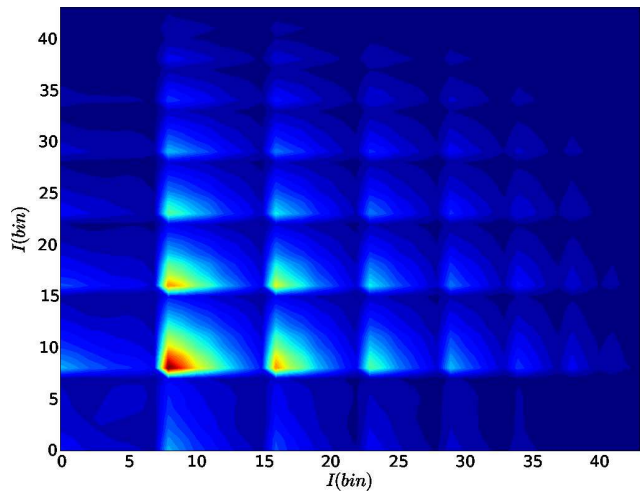}
}
\caption{ This figure illustrates the structure
of our covariance matrix, which follows from
the set up of our data vectors, and the strong
correlations among different bins of the two-
and three-point correlation functions.
}
\label{fig:covmat}
\end{figure}

In order to perform a $\chi^2$ based maximum likelihood analysis,
we estimate a covariance matrix from the 22 mock catalogs available.
These have been extracted from the
Virgo consortium Hubble volume simulation \citep{ColbergEtal2000}.
For details of generating catalogues and biasing
see \citet{ColeEtal1998}. Volume limited subsamples were then
created from each mock in exactly the same way as we do with
the real data of 2dFGRS. Thus, for  every volume limited subsample in
Table~\ref{tb:2dfdata} we have 22 measurements of the two- and
three-point correlation functions. 

We use the usual (biased) estimator for the covariance matrix,
\begin{equation}
  \tilde C_{ij} = \frac{1}{N_{sim}}\sum_{I=1}^{N_{sim}} \Delta d^I_i \Delta d^I_j 
\end{equation}
where $\Delta d^I_i = d^I_i - \avg{d_i}$ 
is the $i$-th element of our data vector (containing
$\xi$ and $\zeta_0$ and filtered according the previous subsection)
from the $I$-th simulation; $N_{sim} = 22$.

It can be shown, that a covariance matrix estimated by the
above equation is necessarily singular if the number of simulations
is less than the length of the data vectors \citep{SzapudiEtal2005}.
More precisely, the rank of the matrix cannot be more than $N_{sim}$.
In practice we have always found this rank to be $N_{sim}-1$ in
several numerical experiments. 

To calculate $\chi^2$ from a singular matrix, we use the singular
value decomposition (SVD) method to create pseudo inverses 
\citep[c.f.][]{GaztanagaScoccimarro2005}. We have decomposed the
$C$ matrix (the tilde denoting estimators is omitted from now on)
as the multiple of three matrices
\begin{equation}
   C = U W V^{T},
\end{equation}
where $W$ is a diagonal matrix, and $U,V$ are orthogonal matrices;
$V^T$ means transpose of $V$. The meaning of this decomposition
is the kernel and image of the linear mapping $C$ 
\citep[c.f.][]{PressEtal1992}.
Moreover, it is similar
to an eigen-vector expansion, unique up to degenerate ``eigenvalues'',
the elements in W. To calculate $\chi^2$, we need $C^{-1}$ of
our singular matrix. Since the inverse is formally $VW^{-1}U^T$,
we can replace entries in $W^{-1}$ corresponding to small
eigenvalues with $0$; this procedure is usually called 
constructing a pseudo-inverse.

\begin{figure}
\resizebox{\hsize}{!}{
\includegraphics{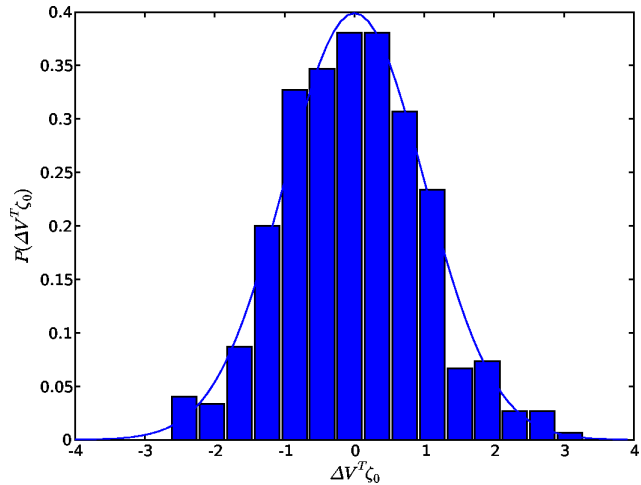}
}
\caption{ The distribution of modes corresponding to
the large eigenvalues in $W$ in our 22 mock surveys.
The mean have been subtracted, and each mode has been
normalized to unit variance. The distribution appears
to be consistent with Gaussian.
}
\label{fig:modes}
\end{figure}

We have found a marked drop in the eigenvalues beyond $N_{sim}-1$. We have
checked that the corresponding columns of $U$ and $V$ are equal,
i.e. they really are ``eigenvectors'' to all intents
and purposes. It is meaningful to calculate $\chi^2$ with the above
pseudo inverse, as long as the distribution of eigenvectors is
Gaussian, as shown in Figure~\ref{fig:modes}. In numerical simulations,
we found that it is safe to keep about half of the eigenvectors,
and for all measurements in this paper we used the top 10 eigenvalues
and their corresponding entries in the $V$ matrix as eigenvectors.
Our results are robust against this choice: have have checked using
$8,9,10,11,12$ eigenvectors, and the results did not change significantly.

\subsection{Maximum Likelihood Analysis}

Using the data vector, and the pseudo inverse of the
covariance matrix defined earlier, we can calculate 
$\chi^2 = \Delta^T C^{-1}\Delta$, as well as likelihoods 
$p\propto \exp(-\chi^2/2)$ , where $\Delta = d - d_{th}(\sigma_8,b,b_2)$,
$d_{th}$ is the theoretical model with explicit dependencies of the
parameters. Note that in principle $C$ should also depend on the
parameters, but we neglect that dependency. This is justified
by the final results, which are not far from the mock surveys,
and simplifies the calculation of likelihood enormously.
To generalize this method, one would need to repeat the
simulations and measurements for each set of parameters, as well
as taking into account the determinant in the likelihood; this
is clearly unfeasible at the moment.

\section{Results}

\begin{table}
\centering
\caption{Our filtering choices for the data vectors
used in the analysis; the lower and upper cut offs
are displayed in units of $\mpc$. 3/P stands for 3 parameter
maximum likelihood analysis based on our phenomenological
model, 2/P and 2/T denotes 2 parameter fits to our
phenomenological and theoretical models, respectively.}
\begin{tabular}{llll}
\hline
$M_{b_J}-5\log_{10} h$ & 3/P & 2/P & 2/T\\
\hline
-18 --- -17 & --          & 11.88 - 34.97 & 18.87 - 34.97\\
-19 --- -18 &  4.04-34.97 & 10.18 - 34.97 & 18.87 - 47.61\\
-20 --- -19 &  4.04-34.97 & 4.04 - 34.97 & 18.87 - 64.8\\
-21 --- -20 &  5.49-34.97 & 11.88 - 34.97 & 18.87 - 64.8\\
-22 --- -21 &  --         & 8.73 - 34.97 & 18.87 - 64.81\\
\hline
\end{tabular}
\label{table:scales}
\end{table}

When applying the above described theoretical framework for the
interpretation of
our clustering measurements, special care needs to be taken about
establishing scale ranges to be included in the data vectors
for maximum likelihood analyses. On the one hand, we want to
include as much data as possible to constrain parameters with
the highest precision, on the other, if we include 
data points where the simple theoretical model is not a good
approximation (for physical reasons and/or because of systematic errors),
we might bias our results. Our considerations are detailed below.

For the theoretical model, a lower cut of $18\mpc$ was used, where
perturbation theory appears to be very accurate. In this
case we use all scales up to $70\mpc$ or the 1/4 of the characteristic
scales of the slice, whichever smaller, to avoid severe edge effects.

For the phenomenological
model we use our measurements from the VLS simulations as theory.
Since we neglect the errors on these measurements, we use an
upper cut of $35\mpc$, above which the errors are non-negligible.
The choice of the lower cut is more delicate. There is a complex
interplay between accuracy of the simple bias and
redshift distortion models we use with the emergence of discreteness
effects. These finally determine the optimal cut in a fairly
subtle way. The apparent complexity motivates an empirical approach:
for the 2-parameter fits we perform maximum likelihood
with a series of low cuts between $4\mpc-14\mpc$ for each
magnitude limit, and finally choose the one with the lowest
$\chi^2$. Fortunately we found that the two-parameter fits are robust 
against this choice, which nevertheless gives the tightest error bars
possible.

For the three-parameter fit the empirical approach would
be too expensive and we also want to maximize the range as much
as possible to resolve the degeneracy between $\sigma_8,b$.
Therefore we use an absolute low cut of $4\mpc$, below which
non-linearities and complexity of the bias are expected to be
strong, complemented with the condition, $\bar n r^3 4\pi/3 \ge 1$.
This choice gives reasonable control over discreteness errors 
which could bias our fit giving a lower cut of $5.1\mpc$ for the $-21 \sim -20$ slice.

In our final choice we took the closest available bin from
our logarithmic binning scheme to the above values. The scale ranges use in
the fits are summarized in Table~\ref{table:scales}.

\begin{figure}
\resizebox{\hsize}{!}{
\includegraphics{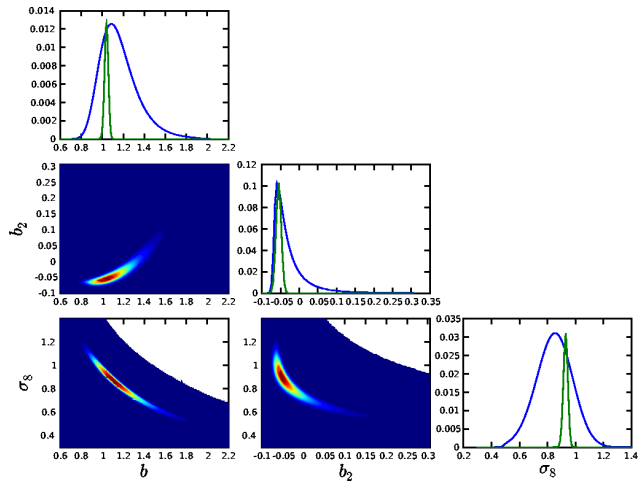}
}
\caption{ Three parameter likelihood contours for the
magnitude slice $-21\sim-20$. The wider (blue) curves correspond
to marginalized distributions, while the narrower (green)
curves are a cut through the likelihood. The cut through the
likelihood curve was normalized to the marginalized distribution
at the maximum. Note that the actual value of the distribution
is somewhat arbitrary, as it depends on the grid size.
We also show
two-dimensional marginalized likelihood contours to 
illustrate the correlations between the parameters.
}
\label{fig:ml3}
\end{figure}

\subsection{Three-parameter Fits}

\begin{table*}
\centering
\caption{Three parameter fits to our two- and three-point clustering
measurements. For each parameter, we present the maximum
of the three-dimensional likelihood with error bars calculated
form a $68$\% of the marginalized distribution. In parentheses
we display the maximum of the marginalized distribution.}
\begin{tabular}{lcccc}
\hline
$M_{b_J}-5\log_{10} h$ & $\sigma_8$ & $b$ & $b_2$ & $\chi^2$\\
\hline
-19 --- -18 & $ 1.07^{+ 0.09}_{- 0.61}$ ( 0.80) & $ 0.81^{+ 0.53}_{- 0.12}$ ( 0.89) & $-0.06^{+ 0.04}_{- 0.03}$  (-0.06) &  0.68 \\
-20 --- -19  & $ 0.90^{+ 0.06}_{- 0.28}$ ( 0.79) & $ 0.97^{+ 0.31}_{- 0.14}$ ( 1.01) & $-0.04^{+ 0.06}_{- 0.02}$  (-0.04) &  3.11 \\
-21 --- -20  & $ 0.93^{+ 0.04}_{- 0.20}$ ( 0.85) & $ 1.04^{+ 0.23}_{- 0.09}$ ( 1.08) & $-0.06^{+ 0.03}_{- 0.01}$  (-0.06) &  0.97 \\
\hline

\end{tabular}
\label{table:ml3}
\end{table*}

We calculated brute force 3 parameter grids with resolution
$0.005$, and 2 parameter grids with resolution $0.001$. 
We checked the effects of grid resolution by repeating
calculations with resolution $0.00025$ without any change
in the results.
While $\sigma_8$ and $b$ are quite degenerate along the line of
$b\times\sigma_8=const.$, as expected, 
the inclusion of triangles
in a large range of scales appears to break the degeneracy,
at least for some of the samples. This is 
evidenced by the fact that a reasonable maximum has developed
for the three volume limited samples in the mid-range. For those,
we present our results in Table~\ref{table:ml3}.

Error bars have been
calculated from the marginalized curves using 68\% thresholds.
In the tables, the overall maximum is quoted first with
error bars, while the maximum of the marginalized likelihood
is presented in parentheses. The $\chi^2$ quoted is normalized
to the degrees of freedom, which in this case is $10-3=7$.

The brightest and faintest magnitude limits do not support
a three-parameter fits. Although they are statistically consistent
with the three others presented, the large error bars due to
the degeneracy of $\sigma_8,b$ makes them meaningless.
Similarly, the theoretical model does not support a stable
three-parameter fit with either of the magnitude slices.

\begin{figure*}
\resizebox{\hsize}{!}{
\includegraphics{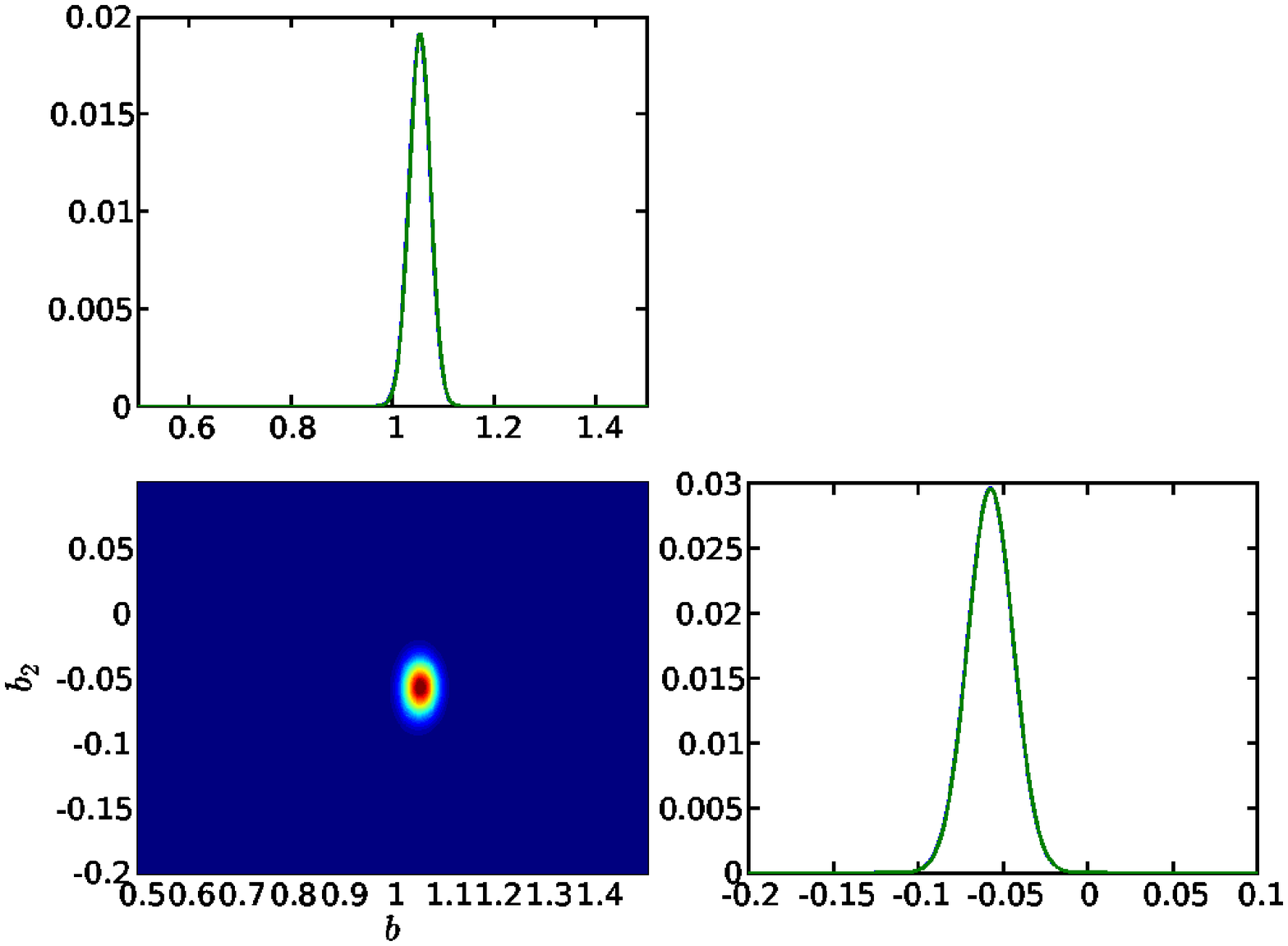}
\includegraphics{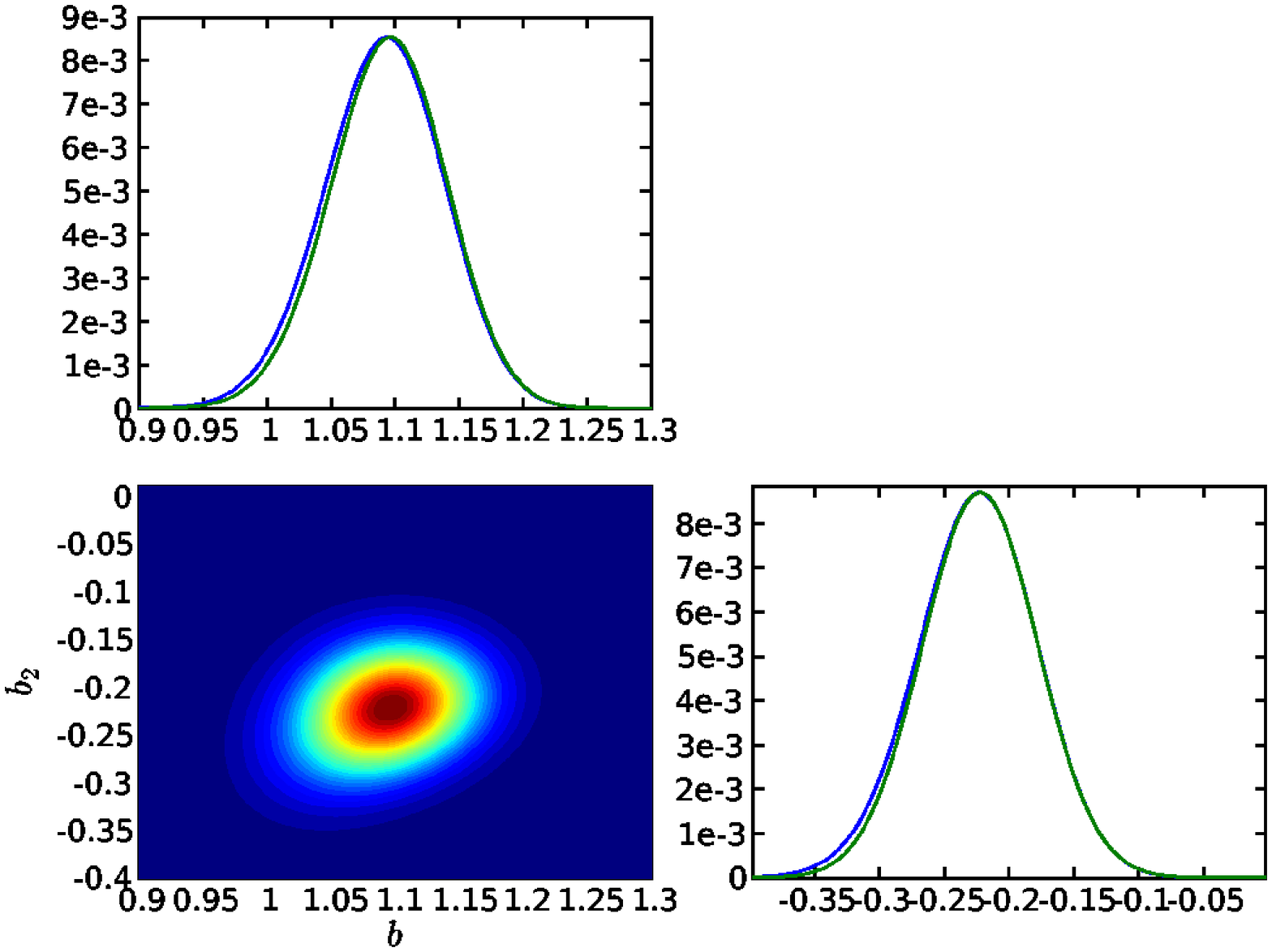}
}
\caption{ Two parameter likelihood contours for the $-21 \sim -20$
slice. Both the phenomenological model (left) and the theoretical
model (right) are displayed.
}
\label{fig:2p2120}
\end{figure*}

\subsection{Two-parameter Fits}

We also performed two-parameter fits using the bias parameters
only. These approximately correspond to fitting 
$b \rightarrow b(\sigma_8/0.9)$. Fixing our reference $\sigma_8=0.9$
pins down the error bars which would otherwise explode
due to the large degree of degeneracy.
$\sigma_8=0.9$ is consistent with all our measurements, and
these fits yield extremely precise values for the bias parameters.
The effective degree of freedom is $10-2=8$.

\begin{table*}
\centering
\caption{Two parameter fit based on the phenomenological model. Notation
is the same as in Table~\ref{table:ml3}}
\begin{tabular}{lccc}
\hline
$M_{b_J}-5\log_{10} h$ & $b$ & $b_2$ & $\chi^2$ \\
\hline
-18 --- -17 & $0.807^{+0.128}_{-0.071}$ (0.833) & $-0.172^{+0.036}_{-0.027}$  (-0.164) & 0.582 \\ 
-19 --- -18 & $0.994^{+0.031}_{-0.036}$ (0.993) & $0.003^{+0.023}_{-0.023}$  (0.003) & 0.343 \\   
-20 --- -19 & $0.969^{+0.027}_{-0.029}$ (0.968) & $-0.035^{+0.011}_{-0.012}$  (-0.035) & 2.721 \\ 
-21 --- -20 & $1.054^{+0.019}_{-0.021}$ (1.053) & $-0.057^{+0.012}_{-0.013}$  (-0.057) & 0.372 \\ 
-22 --- -21 & $1.273^{+0.095}_{-0.070}$ (1.274) & $0.006^{+0.086}_{-0.068}$  (0.010) & 0.206 \\   
\hline
\end{tabular}
\label{table:ml2pz}
\end{table*}

In the case of two-parameter fits, the theoretical model also
supports stable likelihood surfaces and reasonable error bars, 
with the possible exception of the faintest magnitude limit,
which we present for completeness only.

\begin{table*}
\centering
\caption{Two parameter fit based on the theoretical model. Notation
is the same as in Table~\ref{table:ml3}}
\begin{tabular}{lccc}
\hline
$M_{b_J}-5\log_{10} h$ & $b$ & $b_2$ & $\chi^2$ \\
\hline
-18 --- -17  & $0.484^{+0.063}_{-0.322}$ (0.358) & $-0.160^{+0.295}_{-0.334}$  (-0.151) & 1.273 \\	 
-19 --- -18  & $0.955^{+0.087}_{-0.134}$ (0.939) & $-0.252^{+0.132}_{-0.142}$  (-0.256) & 1.248 \\	 
-20 --- -19  & $0.945^{+0.071}_{-0.100}$ (0.935) & $0.185^{+0.053}_{-0.051}$  (0.184) & 2.341 \\	 
-21 --- -20  & $1.096^{+0.043}_{-0.049}$ (1.094) & $-0.222^{+0.044}_{-0.047}$  (-0.223) & 0.747 \\	 
-22 --- -21  & $1.385^{+0.080}_{-0.105}$ (1.375) & $-0.420^{+0.110}_{-0.110}$  (-0.425) & 0.977 \\   
\hline
\end{tabular}

\label{table:ml2t}
\end{table*}

\section{Summary and Discussions}

We presented a measurement of 
the monopole of the three-point correlation function in the 2dFGRS. 
The new technology developed in \citet{Szapudi2004a,PanSzapudi2005}
enabled the estimation of the latter statistics in a wide range of 
scales from $1\mpc-70\mpc$. In the three-point function, up 
to $140\mpc$ scales enter the measurements and the subsequent analysis.
In addition, we measured the two-point correlation function.

\begin{figure}
\resizebox{\hsize}{!}{
\includegraphics{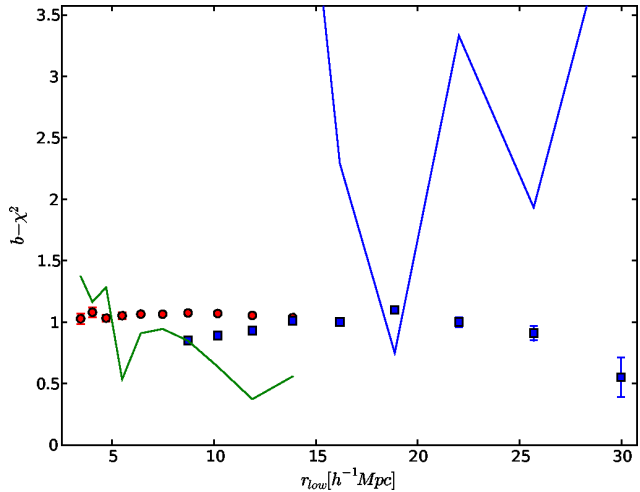}
}
\caption{ We illustrate the degree of robustness of our maximum
likelihood fit by presenting the best fit bias parameters with
error bars against the lower limit used in the fit (the upper limit
was fixed to the value of Table~\ref{table:scales}). The (red) squares
with error bars on the left (scales $4-14\mpc$) show the phenomenological model,
while on the right (blue) is the theoretical model (scales $9-25\mpc$).
The solid lines show the corresponding $\chi^2$'s of the fits.
In general, as long as $\chi^2 \lesssim 1$, the fit is robust.
}
\label{fig:robust}
\end{figure}

To interpret these clustering statistics, we developed a novel 
maximum likelihood
technique based on joint analysis of two- and three-point statistics.
We estimated the joint correlation matrix from mock 2dF surveys.
We seed the top 10 modes detected by an SVD of the joint correlation
matrix to calculate a generalized $\chi^2$. The distribution of
the modes is consistent with Gaussian, therefore we
maximized the Gaussian likelihood with respect the the parameters
of our theory. Our measurements are robust against
keeping more or less eigen-modes, as well as against the scale
range we use for the analysis. In Figure~\ref{fig:robust} we present
the values for the bias parameter $b$ as a function of lower scale
cut for both the phenomenological and theoretical models. Both
models are fairly robust, as long as the $\chi^2 \lesssim 1$.

The significance of our three-parameter fit is that it yields
a highly accurate measurement of $\sigma_8$ from large scale
clustering alone. In particular, our estimate is independent 
of the cosmic microwave background (CMB). Yet, 
our value is in excellent agreement with those
derived from Wilkinson Anisotropy probe
\citep{SpergelEtal2003, FosalbaSzapudi2004}.
$\sigma_8$ is still one of the most uncertain
cosmological parameters, and our technique has a great potential
to further improve the precision of its constraints.
The virtually perfect agreement of $\sigma_8$
from CMB and large scale three-point level clustering is yet another one of
the spectacular successes of the concordance model. In addition,
our result is in excellent agreement with measurements based
on SDSS two-point statistics and joint analysis with WMAP
\citep{TegmarkEtal2004b,PopeEtal2004}.

\begin{figure}
\resizebox{\hsize}{!}{
\includegraphics{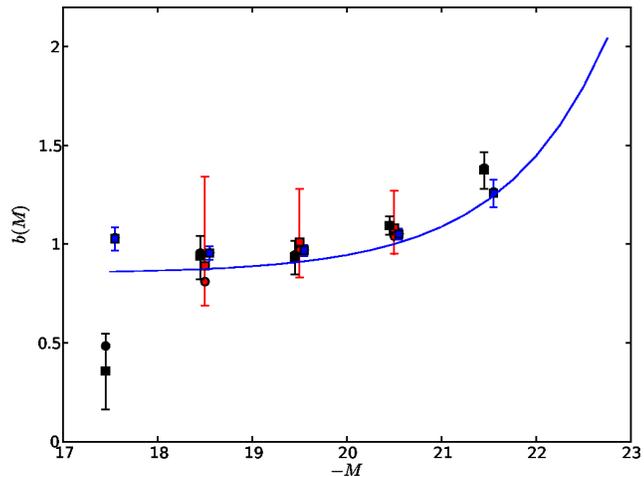}
}
\caption{Comparison of our bias measurements with the
relative bias measurements of \citet{NorbergEtal2002}.
The solid line is $0.85+0.15(L/L^*)$, where we normalized
$L^*$ to be $M = -20.5$ magnitude. Red symbols in
the middle with large error bars are the results of the
three-parameter fit. Blue and black symbols show
the phenomenological and theoretical fits respectively,
and are shifted to left and right slightly for clarity.
Squares and filled circles denote the maximum likelihood
values in two-dimensions and in the marginalized distribution,
respectively.}
\label{fig:bias}
\end{figure}

Our two parameter fits to the bias have extremely small error bars,
they are likely to be systematics (both in the data and in the theory)
limited. They can be considered as a measurement along the
degeneracy line $b \sigma_8/0.9 = $const., and can be directly
compared with previous relative bias measurements from the 2dF.
Figure~\ref{fig:bias} presents a comparison with \citet{NorbergEtal2002},
and shows excellent agreement. 

We see no statistically significant evidence of scale dependency
of the bias. Over the full scale range, from $4\mpc-70\mpc$, 
a constant bias model gave reasonable $\chi^2$'s. 
There is no significant trend between the measurements
on intermediate, and large scales using the phenomenological and
theoretical models respectively: the measured $b$'s are fully
consistent with each other, and with \citet{NorbergEtal2002}.
The only exception is the theoretical model
for the faintest subsample, which also has fairly degenerate
likelihood surfaces. 

A possible check of systematics,
although muddled by cosmic variance, is to estimate the bias
in the NGP and SGP separately. Our phenomenological fits corroborate 
that of \citet{VerdeEtal2002}, who find the SGP slightly more biased:
our estimate is $b  = 1.025^{+0.040}_{-0.048} (1.022)$,
and $b  = 1.097^{+0.020}_{-0.022} (1.096)$ for the NGP and SGP
respectively. Nevertheless, the two samples are consistent 
at the 1-1.5$\sigma$ level, supporting the notion that the
difference between them could be explained with cosmic variance
alone.

The interpretation of our measurements in terms 
of Equations~\ref{eq:model}, \ref{eq:zdist} amounts to be
some of the most comprehensive test of our picture of gravitational
amplification with three point level large scale structure
statistics. In particular, 
numerical experiments and theoretical calculations in the past
tended to focus on a handful of triangular configurations, with
preference for isoceles, and $1:2$ ratios. We have used {\em all}
possible configurations (although monopole only) 
within our dynamic range and logarithmic
binning system, and found a good fit to the data. The simplest
explanation for this is that our basic picture of gravitational
amplification is fundamentally correct. In particular, the
results lend strong support to Gaussian initial conditions,
and thus to inflation,  even though we did not quantify this statement,
since it was a prior in our model.
Figure~\ref{fig:xi-zeta} shows the remarkable success of the
theoretical and phenomenological models down to $1\mpc$. The
most natural interpretation of this is that bias is relatively
simple, and that small scale 
redshift distortions largely cancel non-linear evolution
in redshift space. There is a mild $1.5-2\sigma$ disagreement
between the theoretical model and the data around $30-50\mpc$.
(c.f. inset of Figure~\ref{fig:xi-zeta}). 
While this is not significant according to our measured 
overall $\chi^2$,  it would be interesting to study this
region with more accurate simulations, and higher order theoretical
calculations.

On intermediate scales, we do not detect significant
non-linear bias. On the other hand, the theoretical model on large
scales detected non-linear bias at the $\simeq 2-4\sigma$ level.
This could mean that either the theory is not accurate enough
on the largest scales in accordance to the hint
provided by the inset of Figure~\ref{fig:xi-zeta}, 
and/or the largest scales might have some yet uncovered systematics,
and/or there is significant
non-linear bias. The latter possibility is somewhat unlikely
since the intermediate scales do not display
significant non-linear bias, and it would challenge the well established
notion that bias should become simpler on larger scales. Nevertheless,
\citet{KayoEtal2004} detected a surprizing complexity of non-linear bias with
the three-point correlation function of the Sloan Digital Sky
Survey. To decide between the possible explanations, one would need
a highly accurate measurements of the three-point function 
reliable beyond $35\mpc$. While this can and will be done in 
the future using the Hubble Volume \citep{ColbergEtal2000} or 
similar simulations, the present results do not allow
distinguishing among the above possibilities without the risk of
over-interpreting the data.

The technique we presented for constraining cosmological
and bias parameters from joint likelihood analysis of two-
and three-point statistics has enormous potential for 
further high precision cosmological applications. The constraining
power of the present measurements is limited mainly by the
theory. For one, only 22 simulations have been used to determine
the covariance matrix. More accurate covariance matrix from
a much larger number and realistic mocks could improve the 
statistical power of maximum likelihood estimation based on
the same data. In addition, a more realistic model of bias
and redshift distortions of the three point statistics, perhaps
based on halo models \citep{TakadaJain2003, FosalbaEtal2005}, could
enable the inclusion of all scales measured in the data vector
for even tighter constraints.

The success of the three-parameter fits is remarkable, and it
is a precursor of potentially even more accurate constraints, and
perhaps fits to models with larger number of parameters. 
Perturbation theory can be thought of as a generalized bias
with an anisotropic kernel \citep[c.f.][]{Matsubara1995}.
Since the standard bias model is isotropic, most information
on separating bias from gravitational amplification should
reside in higher order multipoles, such as dipole, and 
quadrupole. Measurements of these in the 2dF will be presented
elsewhere. In particular, the higher order multipoles
contain information on baryonic oscillations \citep{Szapudi2004a},
which in turn might make it possible to constrain further 
cosmological parameters, such as baryon fraction, and dark energy.

Because of the
small number of parameters we used so far, a brute force grid technique was
feasible. If more parameters are fit, 
our technique lends itself naturally to 
Monte Carlo Markov Chain methods \citep[e.g.][]{LewisBridle2002}.
Along the same lines, 
a useful and straightforward follow up to our investigation
is joint likelihood analysis with CMB data. This and other
generalizations will be presented elsewhere.

\section*{Acknowledgement}
IS thanks Alex Szalay for stimulating discussions, and
Zheng Zheng for useful comments.
The authors were supported by NASA through AISR NAG5-11996, 
and ATP NASA NAG5-12101 as well as by
NSF grants AST02-06243, AST-0434413 and ITR 1120201-128440. JP
acknowledges support by PPARC through PPA/G/S/2000/00057.
We acknowledge the heroic effort required to produce, publish,
and keep on-line the 2dFGRS, and sincerly thank all people involved.
The simulations in this paper were carried out by the Virgo
Supercomputing Consortium using computers based at Computing Centre of
the Max-Planck Society in Garching and at the Edinburgh Parallel
Computing Centre. 





\end{document}